\documentclass[a4paper]{jpconf}
\usepackage{graphicx}
\usepackage{amsmath}
\usepackage{amssymb}
\usepackage[hidelinks]{hyperref}
\usepackage{cleveref}

\usepackage[numbers,sort&compress,square]{natbib}

\newcommand{\vv}[1]{\mathbf{#1}}

\crefname{figure}{Fig.}{Figs.}
\Crefname{figure}{Figure}{Figures}
\crefname{equation}{Eq.}{Eqs.}
\Crefname{equation}{Equation}{Equations}
\crefname{section}{Sec.}{Secs.}
\Crefname{section}{Section}{Sections}
\creflabelformat{equation}{#2\textup{#1}#3}

\begin{document}
\title{Assumption Breakdown in Radiative Energy Loss}

\author{Coleridge Faraday}

\address{Department of Physics\char`,{} University of Cape Town\char`,{} Private Bag X3\char`,{} Rondebosch 7701\char`,{} South Africa}

\ead{frdcol002@myuct.ac.za}

\author{W. A. Horowitz}

\address{Department of Physics\char`,{} University of Cape Town\char`,{} Private Bag X3\char`,{} Rondebosch 7701\char`,{} South Africa}

\ead{wa.horowitz@uct.ac.za}

\begin{abstract}
  We show that an integral assumption in DGLV radiative energy loss---the large formation time assumption---is violated at high-$p_T$ for phenomenologically relevant parameters. We further investigate the phenomenological impact of placing a new kinematic bound on the radiated gluon transverse momentum, which ensures that there are no contributions to the energy loss from regions of parameter space that violate the large formation time assumption. We find that this places a large sensitivity on the exact kinematic cutoff used, similar to the known collinear cutoff sensitivity, indicating the theoretical need for a rederivation of DGLV radiative energy with the large formation time assumption relaxed in order to make rigorous predictions. We additionally find that this large formation time cutoff dramatically reduces the size of a short pathlength correction to the DGLV radiative energy loss, which is of phenomenological interest in predicting suppression in small $p +A$ systems. We compute the phenomenological predictions utilising this large formation time cutoff in both $p+A$ and $A+A$ collisions at the LHC, in a convolved radiative and elastic energy loss model.
\end{abstract}

\section{Introduction}

Studying high-$p_T$ particle spectra offers crucial insights into the many-body dynamics of QCD in high-energy collisions. Significant suppression of high-$p_T$ particles in $A+A$ collisions has been observed in RHIC and LHC experiments \cite{PHENIX:2001hpc, STAR:2003pjh, PHENIX:2006ujp, PHENIX:2006mhb}, attributed to parton energy loss in the QGP. This phenomenon aligns well with predictions from pQCD-based models \cite{Dainese:2004te, Schenke:2009gb, Horowitz:2012cf, Wicks:2005gt}. Recent findings in $p+A$ and $p+p$ collisions, including strangeness enhancement \cite{ALICE:2013wgn, ALICE:2015mpp}, quarkonium suppression \cite{ALICE:2016sdt}, and collective behavior \cite{CMS:2015yux, ATLAS:2015hzw}, further support QGP formation. Non-trivial modifications of high-$p_T$ particles have also emerged in small collision systems \cite{ATLAS:2014cpa, PHENIX:2015fgy, ALICE:2017svf}, necessitating theoretical explanations. 

Applying successful $A+A$ models to $p+A$ and $A+A$ collisions presents challenges due to various large system size assumptions. In our previous work \cite{Faraday:2023mmx, Faraday:2023vbo}, we addressed the removal \cite{Kolbe:2015rvk} of the \emph{large pathlength assumption} $L \gg \mu^{-1}$ in the Djordjevic-Gyulassy-Levai-Vitev (DGLV) radiative energy loss model \cite{Gyulassy:2000er, Djordjevic:2003zk}. The correction consists of $\mathcal{O}(e^{- \mu L})$ terms, which were previously assumed to be small, and results in the following novel effects: reduction of energy loss, linear growth with partonic energy, and disproportionate size for incident gluons (cf.\ usual $C_A / C_F$ color factor scaling). This reduction in energy loss could explain the rapid rise of the charged hadron nuclear modification factor with $p_T$ \cite{Horowitz:2011gd} and the enhancement above unity in $\mathrm{p}+\mathrm{A}$ collisions \cite{Balek:2017man,ALICE:2018lyv}.

The derivation of this short pathlength correction, and to a lesser extent the original DGLV derivation, benefited significantly from a \emph{large formation time} assumption---in addition to the usual collinear, Eikonal, and soft assumptions. In \cite{Faraday:2023mmx} we used a numerical energy loss weighted average of dimensionless ratios which are assumed to be small in the derivation of the energy loss single emission kernel, to show that the large formation time assumption was explicitly violated at high-$p_T$ for both the DGLV radiative energy loss and the DGLV radiative energy loss which receives a short pathlength correction.

In this work, we will investigate the phenomenological impact of using a kinematic cutoff on the radiated transverse gluon momentum $\vv{k}$ integral, which ensures that the matrix element (modulus squared) is never evaluated for regions of phase space which violate the large formation time assumption. We will see that this dramatically reduces the size of the short pathlength correction for high-$p_T$ pions in $A+A$
collisions which had previously received a $\sim 100\%$ negative correction at $p_T \sim \mathcal{O}(100)$ GeV, confirming that the short pathlength correction receives a large contribution at high energies from regions of phase space where the large formation time assumption is invalid. We additionally find that both the short pathlength corrected and original DGLV radiative energy loss incur a large sensitivity as a result of this large formation time cutoff. This is not surprising, as there exists a similar sensitivity to the standard collinear cutoff $| \vv{k} |_{\text{max}} = 2x E (1-x)$ \cite{Horowitz:2009eb}, however, we find that the sensitivity to the large formation time cutoff is significantly larger than that of the collinear cutoff.

\section{Discussion and Results}

The radiative energy loss is calculated according to DGLV \cite{Djordjevic:2003zk} with short path length corrections as derived in \cite{Kolbe:2015rvk}.  The number of radiated gluons $N^g$ differential in the momentum fraction radiated away $x$ is given to first order in opacity $L / \lambda$ by
\begin{align}
  &\frac{\mathrm{d} N^g}{\mathrm{d} x} =  \frac{C_R \alpha_s L}{\pi \lambda} \frac{1}{x} \int \frac{\mathrm{d}^2 \mathbf{q}_1}{\pi} \frac{\mu^2}{\left(\mu^2+\mathbf{q}_1^2\right)^2} \int \frac{\mathrm{d}^2 \mathbf{k}}{\pi} \int \mathrm{d} \Delta z \, \rho(\Delta z) \nonumber\\
  &\times\left[-\frac{2\left\{1-\cos \left[\left(\omega_1+\tilde{\omega}_m\right) \Delta z\right]\right\}}{\left(\mathbf{k}-\mathbf{q}_1\right)^2+m_g^2+x^2 M^2}\left[\frac{\left(\mathbf{k}-\mathbf{q}_1\right) \cdot \mathbf{k}}{\mathbf{k}^2+m_g^2+x^2 M^2}-\frac{\left(\mathbf{k}-\mathbf{q}_1\right)^2}{\left(\mathbf{k}-\mathbf{q}_1\right)^2+m_g^2+x^2 M^2}\right] \right. 
   \label{eqn:full_dndx}\\
   &+\frac{1}{2} e^{-\mu_1 \Delta z}\left(\left(\frac{\mathbf{k}}{\mathbf{k}^2+m_g^2+x^2 M^2}\right)^2\left(1-\frac{2 C_R}{C_A}\right)\left\{1-\cos \left[\left(\omega_0+\tilde{\omega}_m\right) \Delta z\right]\right\}\right. \nonumber\\
  &\left.\left.+\frac{\mathbf{k} \cdot\left(\mathbf{k}-\mathbf{q}_1\right)}{\left(\mathbf{k}^2+m_g^2+x^2 M^2\right)\left(\left(\mathbf{k}-\mathbf{q}_1\right)^2+m_g^2+x^2 M^2\right)}\left\{\cos \left[\left(\omega_0+\tilde{\omega}_m\right) \Delta z\right]-\cos \left[\left(\omega_0-\omega_1\right) \Delta z\right]\right\}\right)\right],\nonumber
\end{align}
where the first two lines of the above equation are the original DGLV result \cite{Djordjevic:2003zk} and the last two lines are the correction \cite{Kolbe:2015rvk}. Variable names are the same as in \cite{Djordjevic:2003zk, Kolbe:2015rvk, Faraday:2023mmx}. Importantly, the radiated gluon has 4-momentum (in lightfront coordinates) $k=\left[2 x E, m_g^2+\mathbf{k}^2 /2 x E, \vv{k}\right]$ and the medium exchanged particle $q=\left[q^{+}, q^{-}, \vv{q}\right]$, where $x$ is the radiated (plus) momentum fraction.

The \textit{large formation time assumption} requires that both $\vv{k}^2 / 2 x E \ll \mu_1$ and $(\vv{k} - \vv{q})^2 / 2x E \ll \mu_1$, and is utilised in GLV \cite{Gyulassy:2000er}, DGLV \cite{Djordjevic:2003zk}, and---to a larger extent---in the short pathlength correction to DGLV \cite{Kolbe:2015rvk}. Conventionally, and in our previous work \cite{Djordjevic:2003zk,Kolbe:2015rvk,Faraday:2023mmx}, the radiated gluon, incident parton, and outgoing parton are all enforced to be collinear by setting the upper bound on the $\vv{k}$ integral as $|\vv{k}|_{\text{max}} = 2 x E (1-x)$. Requiring that our matrix element (modulus squared) is not integrated over regions of invalidity for either the large formation time assumption or the collinear assumption, leads to the prescription $|\vv{k}|_{\text{max}} = \text{Min}(\sqrt{2xE\mu_1},2x(1-x)E)$. Note that we enforce only the first of the two large formation time assumptions, as the second assumption defines a far more complicated region of parameter space and we observe numerically that enforcing the first assumption leads to the second holding consistently. We additionally note that at high energies the large formation time cutoff occurs earlier in $\vv{k}$ than the collinear cutoff, which can be understood through the different asymptotic scalings of the cutoffs, $\sim E^{1 / 2}$ vs $\sim E^{1}$ respectively.

\begin{figure}[h]
  \centering
  \includegraphics[width=0.8\textwidth]{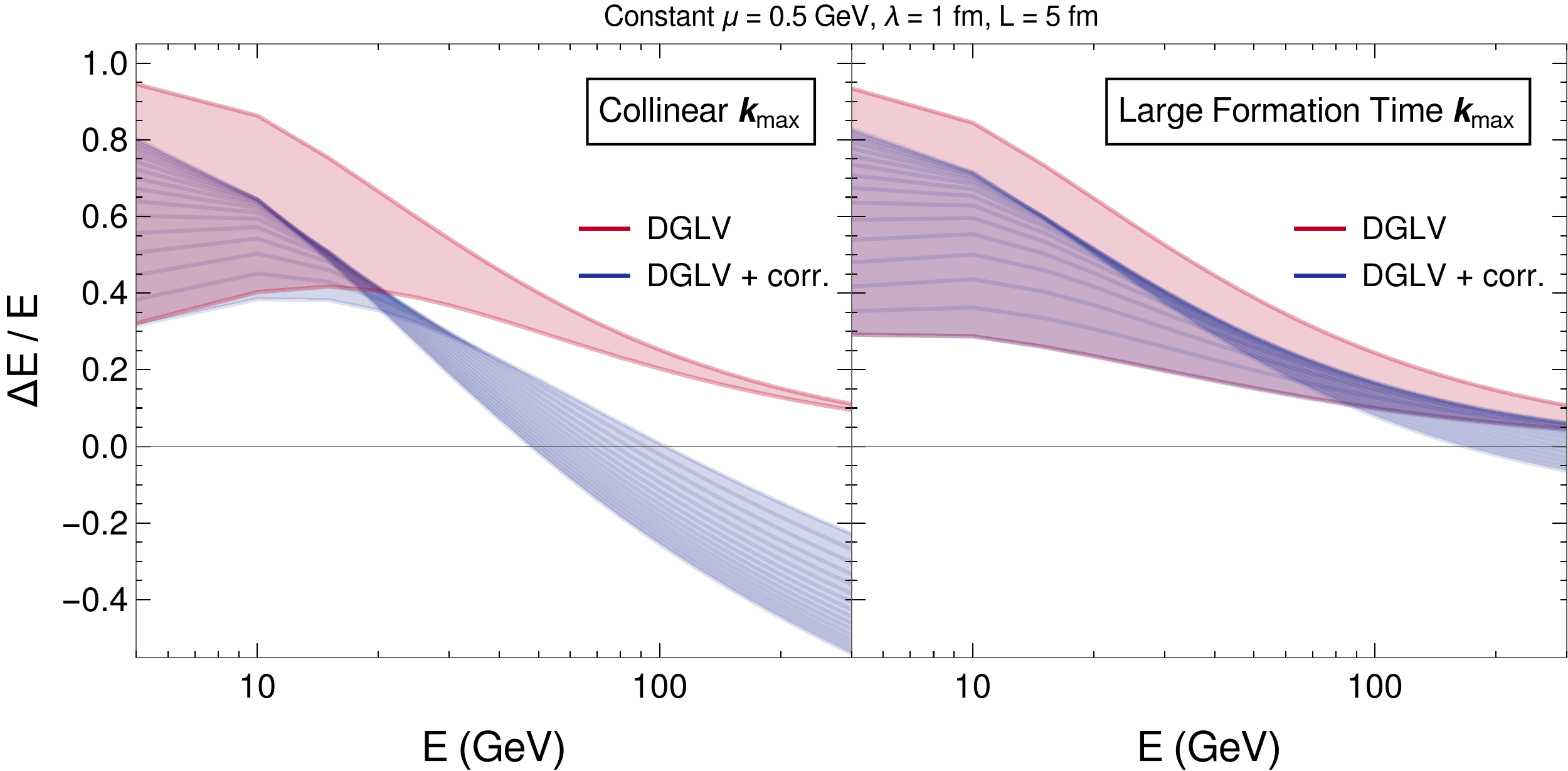}
  \caption{The fractional radiative energy loss calculated according to DGLV and short pathlength corrected DGLV. In the left pane the upper bound on the radiated gluon momentum $\vv{k}$ integral is set by the collinear assumption $| \vv{k}|_{\text{max}} = 2x E (1-x)$, while in the right pane the upper bound is taken as $|\vv{k}|_{\text{max}} = \text{Min}(\sqrt{2xE\mu_1},2x(1-x)E)$. Bands are calculated by varying the upper bound by factors of two. Calculations are done with $\mu = 0.5 GeV$, $\lambda = 1~\mathrm{fm}$, and $L = 5~\mathrm{fm}$.}
  \label{fig:sensitivity}
\end{figure}

A study of the collinear cutoff in the DGLV radiative energy loss was conducted in \cite{Horowitz:2009eb}, where a large sensitivity to the specific choice of cutoff was found. Similarly we show in \cref{fig:sensitivity} the sensitivity of the short pathlength corrected and original DGLV radiative energy loss models to both the large formation time and collinear cutoffs, by varying the respective cutoffs by factors of two. \Cref{fig:sensitivity} shows that by enforcing the large formation time cutoff, one dramatically reduces the size of the short pathlength correction to the radiative energy loss. Further, the uncorrected DGLV radiative energy loss has an increased sensitivity to the specific cutoff used once the large formation time cutoff is imposed (in comparison to only the collinear cutoff), particularly at higher energies $E \gtrsim 100~\mathrm{GeV}$. This sensitivity can be interpreted as the existence of a large weight of the distribution close to the large formation time cutoff. This is further indication that the large formation time assumption breaks down at high $p_T$ for both the short pathlength corrected and original DGLV radiative energy loss, as first discussed in \cite{Faraday:2023mmx}.

In principle one should propagate this theoretical sensitivity to the level of the $R_{AA}$, by computing the $R_{AA}$ which results from the radiative energy loss with $|\vv{k}|_{\text{max}}$---set according to the collinear or collinear and large formation time assumption---scaled by some factor $\alpha \sim \mathcal{O}(1)$. Particularly for the short pathlength corrected energy loss this will be extremely numerically intensive, as the short pathlength corrected energy loss kernel is \emph{not} monotonic in $\mathbf{k}$; meaning that the $R_{AA}$ must be calculated for a full range of $\alpha $. We instead provide a qualitative argument for how the error at the level of $\Delta E / E$ translates to the error at the level of $R_{AA}$

The $R_{AA}$ can be approximated \cite{Faraday:2023mmx} as $R_{AA} \approx \int \mathrm{d} x (1-x)^{n-1} P_{\text{tot}}(x)$ where $P_{\text{tot}}(x)$ is the total probability of a losing a fraction $x$ of the incident parton's energy, and $n$ is the power in the approximately power-law parton production spectra with $n \sim 6$ for gluons at the LHC. If one further assumes that the elastic energy loss is negligible, and take $P_{\text{tot}}(x) = \delta(x - \epsilon)$ where $\epsilon \equiv \Delta E / E = \bar{\epsilon} \pm \Delta \epsilon$, then we find $R_{AA} \sim 1 - (n-1) (\bar{\epsilon} \pm \Delta \epsilon) \sim \bar{R}_{AA} \mp (n-1)\Delta \epsilon$. Under this naive approximation, one sees that the sensitivity to the kinematic $\vv{k}$ cutoff at the level of $\Delta E / E$ are enhanced by $(n-1) \sim 5$ at the level of the $R_{AA}$. Note that this is likely an overestimate of the error as we saw qualitatively in \cite{Faraday:2023mmx} that changes at the level of $\Delta E / E$ are softened by the geometry averaging and Poisson convolution procedures in the full energy loss model.

\begin{figure}[h]
  \centering
  \includegraphics[width=0.47\textwidth]{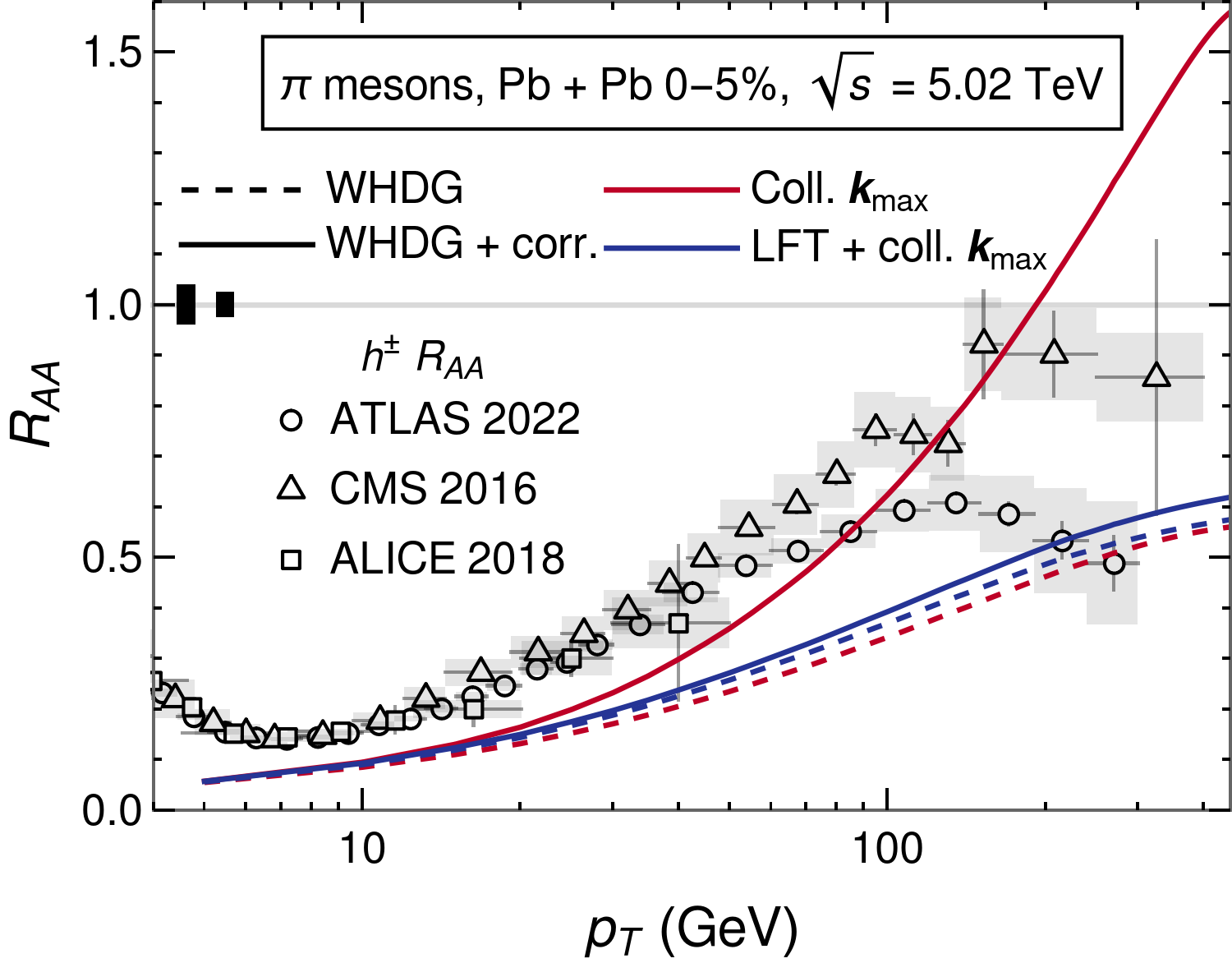}
  \hfill
  \includegraphics[width=0.47\textwidth]{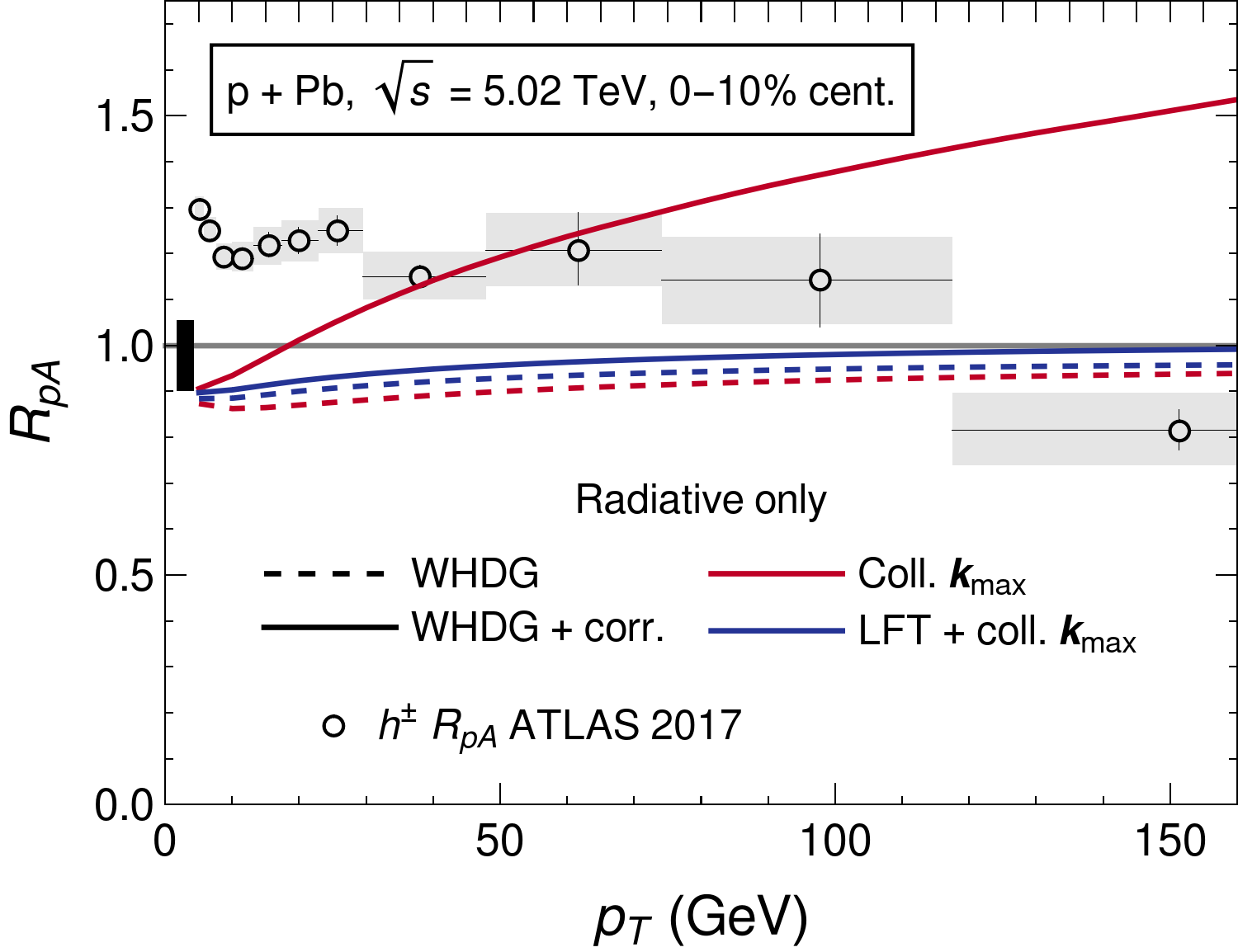}
  \caption{The nuclear modification factor in $A+A$ collisions (left pane) and $p+A$ collisions (right pane) is plotted as a function of the measured transverse momentum $p_T$ for pions. Theoretical predictions are computed for the WHDG model (solid) and the WHDG model which receives a short pathlength correction to the radiative energy loss (dashed). The upper bound of the radiated gluon transverse momentum $\mathbf{k}$ integral is prescribed both by enforcing the collinear assumption (red) and by enforcing both the collinear and large formation time assumptions (blue). In the left pane all plotted curves are computed using only the radiative energy loss and neglecting the elastic energy loss. Data are from ATLAS \cite{ATLAS:2022kqu,Balek:2017man}, CMS \cite{CMS:2016xef}, ALICE \cite{Sekihata:2018lwz}}.
  \label{fig:raa}
\end{figure}

The phenomenological implications of such a large formation time cutoff are investigated using the convolved radiative and elastic energy loss model described in \cite{Faraday:2023mmx}, based on the Wicks-Horowitz-Djordjevic-Gyulassy (WHDG) model. The left pane of \cref{fig:raa} shows the predicted $R_{AA}$ for pions in central $\mathrm{Pb} + \mathrm{Pb}$ collisions as a function of final transverse momentum $p_T$ versus data. Including the large formation time cutoff dramatically reduces the size of the correction, as is expected from \cref{fig:sensitivity}. The effect on the $R_{AA}$ calculated with the uncorrected WHDG is negligible. We note however that \cref{fig:sensitivity} indicates that there will likely be a large sensitivity at the level of the $R_{AA}$ to the exact kinematic cutoff on the $\mathbf{k}$ integral used.

The right pane of \cref{fig:raa} shows a similarly dramatic decrease of the size of the short pathlength correction once the large formation time cutoff is enforced, as well as the same negligible effect on the uncorrected DGLV energy loss. Note that in \cref{fig:raa} we only include the radiative energy loss since, as was previously shown in \cite{Faraday:2023mmx,Faraday:2023vbo}, the central limit theorem approximation to the elastic energy loss breaks down in small systems. In the future, this will need to be corrected by the removal of the central limit theorem approximation in the elastic energy loss as well as potential short pathlength corrections to the elastic energy loss.

\section{Conclusion}

We provided an argument for the breakdown of the large formation time assumption in both DGLV \cite{Djordjevic:2003zk} and the short pathlength correction to DGLV \cite{Kolbe:2015rvk}. We showed that one can artificially impose a cutoff on the radiated gluon momentum $\mathbf{k}$ integral which ensures that the matrix element (modulus squared) is not integrated over regions where the large formation time assumption is invalid. Doing so dramatically reduces the size of the short pathlength corrected DGLV radiative energy loss, indicating that the short pathlength correction to the DGLV radiative energy loss receives an increasingly large contribution from regions where the large formation time assumption is invalid at high $p_T$. The effect on the uncorrected DGLV radiative energy loss is small $\sim 10\%$. We explore the sensitivity of the energy loss to the exact kinematic bound imposed, by varying kinematic cutoff by factors of two. Doing this we find that imposing the large formation time cutoff, as opposed to the usual collinear cutoff, increases the sensitivity of both the DGLV and short pathlength corrected DGLV to the exact kinematic cutoff chosen.

We further calculate both the $R_{AA}$ and $R_{pA}$ for pions produced in central collisions at the LHC. We find that the dramatic decrease of the correction at the level of $\Delta E / E$ translates to a dramatic decrease in the predicted $R_{AA}$ and $R_{pA}$ as is expected. We note that the sensitivity incurred on the $\Delta E / E$ will likely translate to a similar sensitivity on the $R_{AA}$ and $R_{pA}$. We argue therefore that GLV, DGLV, and the short pathlength correction to DGLV all need to be rederived, with the large formation time assumption relaxed, for rigorous quantitative predictions to be made for \emph{either} $A+A$ or $p+A$ collisions.

Possible future investigations into small systems could involve establishing a more robust foundation for energy loss calculations \cite{Clayton:2021uuv}, exploring adjustments to thermodynamics due to system size \cite{Mogliacci:2018oea}, analyzing the equation of state in small systems \cite{Horowitz:2021dmr}, and calculating small system size corrections to the effective coupling \cite{Horowitz:2022rpp}.

\section*{Acknowledgements}

CF and WAH thank the National Research Foundation and the SA-CERN collaboration for their support.

\providecommand{\newblock}{}

\end{document}